# Experimentally Engineering the Edge Termination of Graphene Nanoribbons

Xiaowei Zhang[1,2†], Oleg V. Yazyev[1,2,3†], Juanjuan Feng[1,4†], Liming Xie[5†], Chenggang Tao[1,2], Yen-Chia Chen[1,2], Liying Jiao[5], Zahra Pedramrazi[1], Alex Zettl[1,2], Steven G. Louie[1,2], Hongjie Dai[5], and Michael F. Crommie[1,2] *

[1] Department of Physics, University of California at Berkeley, Berkeley, California 94720, USA

[2] Materials Science Division, Lawrence Berkeley National Laboratory, Berkeley, California 94720, USA

[3] Institute of Theoretical Physics, Ecole Polytechnique Fédérale de Lausanne (EPFL), CH-1015 Lausanne, Switzerland

[4] School of Physical Science and Technology, Lanzhou University, Lanzhou, Gansu 730000, China

[5] Department of Chemistry and Laboratory for Advanced Materials, Stanford University, Stanford, California 94305, USA

† These authors contributed equally to this work.

* Corresponding author. E-mail: crommie@berkeley.edu

# Abstract

The edges of graphene nanoribbons (GNRs) have attracted much interest due to their potentially strong influence on GNR electronic and magnetic properties. Here we report the ability to engineer the microscopic edge termination of high quality GNRs *via* hydrogen plasma etching. Using a combination of high-resolution scanning tunneling microscopy and first-principles calculations, we have determined the exact atomic structure of plasma-etched GNR edges and established the chemical nature of terminating functional groups for zigzag, armchair and chiral edge orientations. We find that the edges of hydrogen-plasma-etched GNRs are generally flat, free of structural reconstructions and are terminated by hydrogen atoms with no rehybridization of the outermost carbon edge atoms. Both zigzag and chiral edges show the presence of edge states.

**Keywords**: graphene nanoribbon, synthesis, scanning tunneling microscopy, first-principles calculations

The edges of graphene exhibit several unique features, such as the presence of localized edge states, and are anticipated to provide an important means of controlling the electronic properties of this two-dimensional material.[1-3] In particular, edges oriented along the high-symmetry zigzag direction or along any low-symmetry (chiral) direction give rise to unique localized edge states[4-9] that are predicted to result in magnetic ordering.[1-3] Such edge-dependent behavior is expected to be even more pronounced in ultra-narrow strips of graphene, dubbed nanoribbons, where edges make up an appreciable fraction of the total nanostructure volume, thus creating new nanotechnology opportunities regarding novel electronic and magnetic nanodevices.[2,3] Edge states in chiral nanoribbons have been experimentally observed,[10,11] but it has so far not been possible to control and correlate nanoribbon edge electronic structure with specific chemically defined terminal edge groups.

Here we report a scanning tunneling microscopy (STM) study of graphene nanoribbons (GNRs) that are treated by hydrogen plasma etching. We find that hydrogen plasma etches away the original edge groups and develops segments with different chiralities along the edge. We have closely examined three different types of representative GNR edge segments: zigzag segments, (2,1) chiral edge segments, and armchair segments. Comparison between our experimental data and first-principles simulation of energetically most favorable structures shows good agreement. For example, we find that the edge carbon atoms of our etched GNRs are terminated by only one hydrogen atom, and that both zigzag and chiral edges show the presence of edge states. The edges of hydrogen-plasma-etched GNRs are seen to be generally free of structural reconstructions and curvature[10,11], with the outermost carbon edge atoms being in the *sp$^2$* hybridization state. Hydrogen plasma etching thus enables the engineering of GNR edges from an unknown terminal group (with associated edge curvature[10,11]) to a flat edge morphology with known atomic termination and terminal bonding symmetry.

**RESULTS AND DISCUSSION**

The investigated GNRs were obtained by hydrogen plasma treatment[12] of chemically unzipped carbon nanotubes[13] deposited onto a Au(111) (see the Methods section). Prior to hydrogen plasma treatment, these GNRs typically exhibit curved edges[10,11,14] that hinder access *via* STM to the very outermost edge atoms. The chemical nature of the pre-etched outermost atoms is therefore unknown, but based on the GNR chemical treatment they are likely terminated with some form of oxygen-containing functional groups.[10] Figure 1a shows a typical room temperature STM image of a GNR that has been deposited onto Au(111) before being etched by hydrogen plasma. The line profile indicates typical edge curvature, where the curved part of the edge has a width of 5 nm and a height of 0.3 nm above the center terrace region of the GNR.

The effect of hydrogen plasma treatment on these GNRs is two-fold. First, the hydrogen plasma etches away the original edge groups, and substitutes them with hydrogen (the simplest possible monovalent edge termination). Second, the edges become significantly rougher and develop short segments (several nanometers long) that display different chiralities within the same GNR (Figures 1b, 2a-e) (the entire GNR thus does not achieve global thermodynamic equilibrium that would result in an overall preferred edge orientation). The combination of these two factors changes the interaction between the edges and the substrate, resulting in a flat, uncurved morphology with the outermost edge atoms being more exposed. Figure 1b shows that the bright strips due to edge curvature are no longer visible in etched GNRs. Instead, the etched GNRs are flat, with a height similar to the height of the interior terrace of unetched GNRs, indicating that the etching process starts from the edges and moves towards the center.

Higher resolution topographic images (Figures 2a-e) on different parts of an etched GNR show the honeycomb structure of the interior graphene. By superimposing a hexagonal lattice structure, we are able to identify the chirality of each segment of the GNR edge (see Supplemental material). Figures 2c-e show close-up images of three different types of representative GNR edge segments: a zigzag segment, a chiral edge segment orientated along the (2,1) vector of the graphene lattice, and an armchair segment, respectively. The 2-nm-long zigzag edge segment (Fig. 2c) appears as a sequence of bright spots visible along the edge, which then decays into the interior graphene. This segment exhibits a small depression near the middle of the outer row of edge atoms, while the second row of edge atoms next to the depression appear to be brighter than adjacent second row atoms. The (2,1) chiral edge segment (Fig. 2d) shows a periodic modulation in STM intensity along its length. Comparison with a superimposed lattice structure (see Supplementary material) shows that the periodic bright spots are localized along zigzag-like fragments. A break in the periodic pattern is observed in the middle of the chiral edge, possibly due to the presence of a vacancy defect. The armchair edge (Fig. 2e), in contrast, shows no edge enhancement in the STM intensity. Instead, the armchair edge exhibits a pronounced standing-wave feature with periodicity of ~0.4 nm at the −0.97 V bias voltage used in our measurements.

Because the experiments were carried out at room temperature (see Methods section) thermal effects limit our ability to perform highly resolved STM spectroscopy. Nevertheless, our STM images contain a significant amount of information regarding GNR edge electronic structure. In order to understand this information we must first determine the bonding arrangement of hydrogen atoms at the GNR edges. Once we know this bonding arrangement we can then calculate the GNR electronic local density of states and compare it to the STM data to self-consistently confirm the structural model and electronic behavior. Our strategy for performing this procedure is to first calculate the energetic stability of different edge structures, and then to use the thermodynamically favorable structures to guide our first-principles electronic property calculations which are then compared to experiment.

We determined the thermodynamically most favorable structures by calculating the edge formation energy of different hydrogen-bonded GNR edge structures in contact with a reservoir of hydrogen. Edge carbon atoms were allowed to terminate with either one ($sp^2$ hybridization) or two ($sp^3$ hybridization) hydrogen atoms, and we restricted our consideration only to structural terminations having the same periodicity as the unterminated bare edge, since such symmetry is observed experimentally. Different hydrogenated edge structures differ in their local chemical composition, and thus their formation energies per edge unit length depend on the chemical potential of hydrogen, $\mu_\text{H}$, according to[15]

$$G(\mu_\text{H}) = \frac{1}{2a}\left(E_\text{GNR} - \frac{N_\text{C}}{2}E_\text{graphene} - \frac{N_\text{H}}{2}E_{\text{H}_2} - N_\text{H}\mu_\text{H}\right),$$

where $a$ is the edge periodicity, $N_C$ and $N_H$ are the number of carbon and hydrogen atoms per unit cell, and $E_{GNR}$ and $E_{graphene}$ are the total energies of the model GNR and ideal graphene per unit cell, respectively. The chemical potential, $\mu_H$, here defined using the total energy $E_{H_2}$ of an $H_2$ molecule as a reference, is a free parameter which depends on particular experimental conditions. For this reason, we analyzed a broad range of chemical potentials, as shown in Figure 3. Structures having the lowest formation energies, $G(\mu_H)$, at a given $\mu_H$ are highlighted with thick lines and the corresponding structures are shown below with the $\pi$ bonding network emphasized. We note that more stable structures with long-range periodicity can in principle be realized,[16] but they are not observed here since global thermodynamic equilibrium is not achieved under the present experimental conditions.

For the zigzag edge (Fig. 3a), two hydrogen configurations are possible – either a "simple" $sp^2$-bonded hydrogen zigzag edge for $\mu_H < 0.33$ eV or an $sp^3$-bonded edge with two hydrogen atoms per carbon edge atom for $\mu_H > 0.33$ eV (the latter results in the Klein edge $\pi$ bonding network topology[17]). The shaded region in Figure 3 shows the condition for graphene to transform into graphane [18,19] with a full basal hydrogenation of stoichiometry CH ($\mu_H > -0.2$ eV). Since this is not observed experimentally, meaning $\mu_H < -0.2$ eV, we are able to exclude the Klein edge scenario. We thus conclude that the zigzag GNR is terminated with one hydrogen atom per carbon edge atom.

The armchair edge (which has two carbon edge atoms per unit cell) can, in principle, support three possible hydrogen terminations. As shown in Figure 3b, however, only two of them have regions of stability – either both carbon edge atoms terminated with one hydrogen atom ($\mu_H < -0.19$ eV) or both carbon edge atoms terminated with two hydrogen atoms ($\mu_H > -0.19$ eV). These two configurations are equivalent from the point of view of the $\pi$ electron system topology, and both have an identical electronic structure that is consistent with the experimental observation. However, the condition of observing graphene instead of graphane requires $\mu_H < -0.2$ eV, thus indicating that the armchair edge is most likely terminated with one hydrogen atom per carbon atom.

The situation is somewhat more complicated for the case of the (2,1) chiral edge, which has three inequivalent edge atoms per unit cell and thus can realize, in principle, eight distinct possible hydrogen terminations. Only three of these, however, have regions of stability (Fig. 3c): the case where all edge atoms are terminated with one hydrogen atom ($\mu_H < -0.19$ eV), the case where two adjacent edge atoms are terminated with two hydrogen atoms and the third edge atom is bonded only to one hydrogen ($-0.19$ eV $< \mu_H <$ 0.33 eV), and the case where all three edge atoms are each terminated with two hydrogen atoms ($\mu_H >$ 0.33 eV). The first two structures realize the same $\pi$ electron network topology. However, because the upper limit of $\mu_H$ realized under the present experimental condition is $-0.2$ eV, we conclude that the (2,1) chiral edge termination should involve only one hydrogen atom per edge carbon atom. This chiral edge termination is equivalent to the previously used models of Refs. 10 and 20.

To further confirm that the calculated thermodynamically favorable edge terminations correspond to what we observe experimentally, we performed first-principles simulations (see Methods section) of the STM images for these structures (Figures 2f-h) and compared them to our experimental data. The structural models were based on the energetically most stable termination (i.e. one hydrogen atom per edge carbon atom), and did not involve any covalent bonding reconstructions other than six-membered rings. We did not include the Au(111) substrate in our calculations since Au does not have a significant effect on the overall electronic structure of graphene.[21]

The resulting simulated STM images nicely match the experimental data. This can be seen first for the zigzag segment in Figure 2f, which shows a sequence of bright spots along the edge. A single carbon atom was removed from the center of the zigzag edge in the simulation, and this is seen to explain the depression in the middle of the outer row of atoms and the slight enhancement in the second row of carbon atoms. The simulation of the (2,1) chiral edge (Fig. 2g) shows very pronounced edge states in agreement with the experimental data. It also features the observed intensity modulation along the length of this edge, which results from edge states localized along the zigzag-like fragments. An extra pair of edge carbon atoms was added to the middle of this edge segment which effectively elongates one of the zigzag-like fragments and shortens the neighboring one, thus explaining a break in the periodic pattern observed in the middle of the experimental edge segment. The simulated armchair segment (Fig. 2h) does not show intensity enhancement along the edge, as seen experimentally for this structure. This is consistent with the fact that this edge orientation does not give rise to localized states (this is further confirmed by average linescans shown in the Supplementary material). The armchair edge simulation also features a standing-wave pattern, in agreement with previously reported predictions[22,23] and the experimental data of Figure 2e. Other possible edge terminations were also simulated for these different edge orientations (see Supplementary material), but they did not reproduce the experimental results nearly as well as those shown. This in-depth experiment/theory comparison provides confirmation that the experimentally observed edge terminations match the theoretically derived most stable hydrogen configurations.

**CONCLUSIONS**

We have investigated hydrogen plasma treated GNRs on a Au(111) substrate. We find that hydrogen plasma etches away unknown edge terminal groups and promotes formation of short segments having different chiralities along the edge. From more than 20 GNRs examined in this study, we observed no apparent preferred orientation of the edge segments. The chirality of edge segments likely has some dependence on the initial chirality of the whole GNR, local environments, and the out-of-equilibrium nature of the hydrogen plasma etching; thus thermodynamics is expected to play only a minor role in determining the overall statistics of edge orientations. We have primarily studied three types of GNR edge segments: zigzag segments, (2,1) chiral edge segments, and armchair segments. Our combination of local probe microscopy and *ab-initio* simulations enables us to determine both the terminal hydrogen-bonding structure and the edge electronic structure for edge-engineered graphene nanoribbons. This work has important implications for graphene research and technology as it introduces a new method for controlling the chemical termination and direction of GNR edges required for manipulating their electronic properties.

**METHODS**

**Experiments.** The experiments were performed using an Omicron VT-STM operating at room temperature. The Au(111) substrate was cleaned by standard sputter-annealing procedures in ultra-high-vacuum (UHV) before being transferred *ex situ* for spin-coating of GNRs. The GNRs were chemically fabricated using carbon nanotube unzipping methods (as in Ref. 13). The sample was then exposed to hydrogen plasma for 15 minutes (following the procedures described in Ref. 12). The sample was then placed back into the UHV chamber where it was annealed up to 500 °C for several hours before being transferred *in situ* for UHV STM measurements.

**Calculations.** First-principles electronic structure calculations were carried out using the local spin density approximation of density functional theory. The model for the first-principles simulations of STM images are approximately 7-nm-wide graphene nanoribbons with hydrogen-terminated edges. These large-scale simulations of STM images have been performed using the SIESTA package[24] and a combination of a double-$\zeta$ plus polarization basis set, norm-conserving pseudopotentials,[25] and a mesh cutoff of 200 Ry. The atomic positions have been fully relaxed. The STM intensities were calculated using the the Tersoff-Hamann approximation[26] assuming a fixed tip sample distance of 5 a.u. and a negative bias of −0.97 V in accordance with experimental conditions. The stabilities of various edge terminations were investigated using moderate size GNR models of approximately 1.5 nm width and a plane-wave-based computational scheme implemented in the Quantum-ESPRESSO package.[27] In these calculations we used a combination of ultrasoft pseudopotentials,[28] and plane-wave kinetic energy cutoffs of 30 Ry and 300 Ry for wavefunctions and charge density, respectively.

*Acknowledgements:* Research supported by the Office of Naval Research Multidisciplinary University Research Initiative (MURI) award no. N00014-09-1-1066 (GNR sample preparation and characterization, thermodynamic stability calculation), by the Director, Office of Science, Office of Basic Energy Sciences of the US Department of Energy under contract no. DE-AC02-05CH11231 (STM instrumentation development and development of numerical simulation tools), and by National Science Foundation award no. DMR10-1006184 (calculation of theoretical LDOS). O. V. Y. acknowledges support from the Swiss National Science Foundation Grant No. PP00P2_133552 (development of new techniques for edge simulation). L. X., L. J. and H. D. acknowledge support from Samsung and MARCO MSD (development of nanoribbons synthesis and plasma etching).

*Supporting Information Available:* Determination of chiralities of GNR segments and location of edge states, simulation of different hydrogen termination for zigzag and (2,1) chiral edges, and comparison of average linescan profiles between experimental images and simulations. This material is available free of charge *via* the Internet at http://pubs.acs.org.

**Figure Captions**

Figure 1: **Effect of hydrogen plasma treatment on GNRs deposited on a Au(111) substrate.** (a) Room temperature constant-current STM topograph ($V_S$ = 1.5 V, $I_t$ = 100 pA) of a GNR before hydrogen plasma etching. (b) Room temperature STM image of a GNR after hydrogen plasma treatment ($V_S$ = −1.97 V, $I_t$ = 50 pA). Insets show the indicated line profiles.

Figure 2: **Atomically-resolved STM topographs of GNR edges: experiment vs. first-principles simulations.** (a, b) Larger scale room temperature STM topographs of two segments of a GNR ($V_S$ = −0.97 V, $I_t$ = 50 pA). (c, d, e) Zoomed-in atomically-resolved images of edge segments circled in figs. 2a,b having different chiralities: a zigzag edge ($V_S$ = −0.97 V, $I_t$ = 50), a (2, 1) chiral edge ($V_S$ = −0.97 V, $I_t$ = 50 pA), and an armchair edge ($V_S$ = −0.97 V, $I_t$ = 50 pA), respectively. (f, g, h) STM images simulated from first principles using the Tersoff-Hamann approximation for the STM tunneling current, the same bias voltage as in the experiments, and the thermodynamically most stable hydrogen edge configuration. These simulations suggest that the plasma treatment results in simple edge termination with one hydrogen atom saturating each dangling bond. The atomic structures of the underlying lattices of carbon atoms are shown as black lines.

Figure 3: **Thermodynamic stability of hydrogenated graphene edges calculated from first principles.** Edge formation energy per unit length ($G$) as a function of chemical potential of hydrogen ($\mu_H$) calculated from first principles for various hydrogen termination patterns for (a) a zigzag, (b) an armchair, and (c) a (2,1) chiral edge of a GNR. The structures for stable edge terminations are sketched below. The $sp^2$ carbon atom bonding networks are highlighted in color (matched to the energy plot) while $sp^3$

carbon atoms and terminating bonds are shown in black. The shaded areas denote the range of chemical potentials $\mu_H$ for which graphane is more stable than graphene.

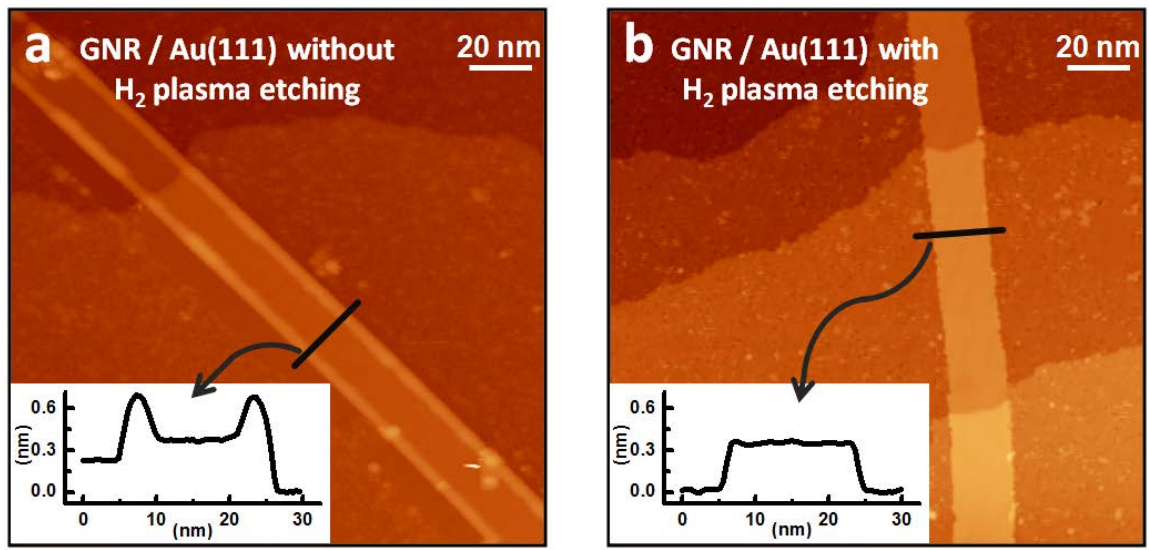

**Figure 1**

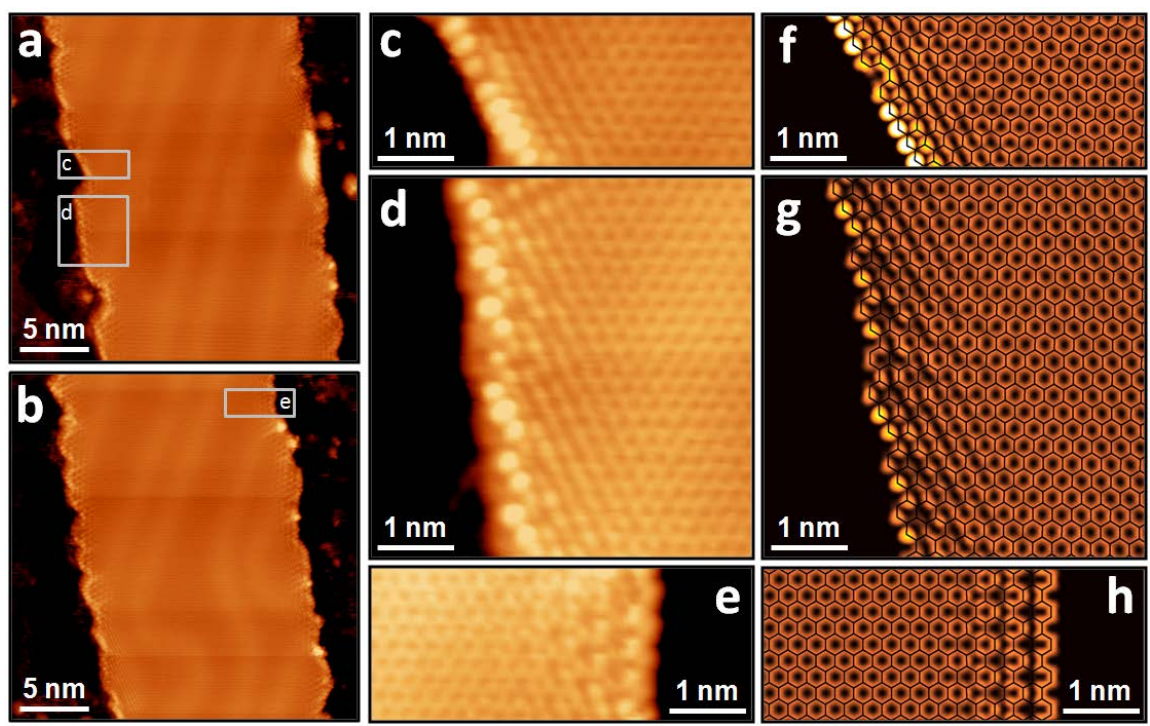

**Figure 2**

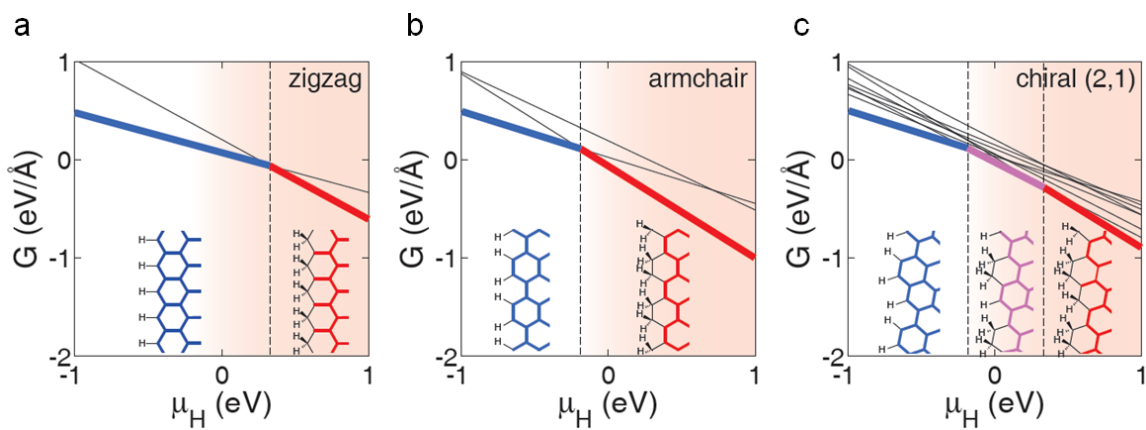

**Figure 3**

# Supplementary Information for "Experimentally Engineering the Edge Termination of Graphene Nanoribbons"

**Contents**

1. Determination of chiralities of GNR segments and location of edge states
2. Simulations of different hydrogen terminations for zigzag GNR edge
3. Simulations of different hydrogen terminations for the (2, 1) chiral edge
4. Comparing average linescan profiles between experimental images and simulations

**1. Determination of chiralities of GNR segments and location of edge states**

The flat morphology of the etched GNRs and the atomically resolved STM images allow us to unambiguously determine the chirality of each segment. In Figure S1, two dashed black lines lie parallel to the edge orientation and the zigzag orientation respectively. The chiral angle between them is 19.1°, from which we determine that this segment is along the (2,1) vector of the graphene lattice. Superimposing the graphene lattice structure with the (2,1) edge orientation reveals that the STM intensity enhancement is localized along the zigzag-like fragments.

**2. Simulations of different hydrogen terminations for zigzag GNR edge**

For the zigzag GNR, there are two basic hydrogen terminations – with either one or two hydrogen atoms terminating the outermost carbon atoms. In the first case, the terminated carbon atoms have $sp^2$ hybridization and thus contribute to the π-electron network of graphene. When terminated with two hydrogen atoms, the edge carbon atoms assume $sp^3$ hybridization and do not contribute to the π-electron system. This configuration has the π-electron network topology of the so-called Klein edge. Both $sp^2$ and Klein edge terminations give rise to edge states localized on only one of the sublattices of graphene, but in each case the sublattice is different. Simulated STM images for these two cases (calculated for the same bias voltage as in the experiment) are shown in Figures S2a ($sp^2$ case) and S2b ($sp^3$ case). These simulated images show edge states localized on different sublattices of graphene, thus allowing the two cases ($sp^2$ versus $sp^3$ hydrogen bonding) to be distinguished through comparison with experimental images. The corresponding experimental image for a zigzag edge with superimposed lattice structure is shown in Figure S2c. By observing which sublattice the experimental intensity enhancement is associated with it is possible to determine that this is the $sp^2$-bonded case and not the $sp^3$-bonded case. This provides further evidence (beyond our calculations of thermodynamic stability) that the experimentally observed zigzag edge has only one hydrogen atom per edge carbon atom.

3. **Simulations of different hydrogen terminations for the (2,1) chiral edge.**

The (2,1) chiral edge has 3 inequivalent positions of edge carbon atoms (see Figure S3). Thus, there are possible $2^3 = 8$ different configurations in which either 1 or 2 hydrogen atoms terminate each edge carbon atom. The simulated STM images of all 8 configurations are shown in Figure S3a – h. Only three of these configurations (Figures S3a, g, h) have regions of stability as shown in Figure 3c of the main text. Only two of them, the normal chiral edge with one hydrogen atom per edge carbon atom (Figure S3a) and the one with two hydrogen atoms terminating edge carbon atoms in position 2 and 3 (Figure S3g) qualitatively agree with the experimental STM image (see Figure 2d). These two cases are electronically equivalent since they share the same π-electron system boundary. However, the structure with two hydrogen atoms per carbon atom lies in the regime where graphene is thermodynamically less stable than graphane (see main text), and so we conclude that the observed termination of the (2,1) chiral edge has one hydrogen atom per edge carbon atom.

4. **Comparing average linescan profiles between experimental images and simulation**

Here we further compare the experimental data and simulations for GNR edge electronic structure by examining average line profiles perpendicular to zigzag and armchair. We took more than 20 parallel line scans from the experimental data in the shaded regions of Figures S4a and S4b, and then averaged them to get the blue curves in Figures S4e and S4f. For the simulation images, we first used a mean-filtering image processing method to account for the finite size of the STM tip, and we then took an average of parallel line scans oriented perpendicular to the edges. The theoretical line scans obtained in this way are depicted as red dashed lines in Figures S4e and S4f, and are offset vertically for clarity.

For the zigzag edge, both the experimental and theoretical line scans exhibit an LDOS oscillation with a period of 2.1 Å, which is close to the distance between neighboring zigzag chains. This oscillation can be explained by the fact that the localized edge state decays exponentially over zigzag chains away from the edge. For the armchair edge, a different modulation period of 3.8 Å is seen. This can be explained by intervalley scattering of electrons.[23] The zigzag edge is seen to have a large buildup in LDOS near the edge (in both the experiment and the simulation) which is not seen for the armchair edge. This is due to the fact that the zigzag edge has an edge state while the armchair edge does not.

**Figure Captions.**

**Figure S1: Determination of chiralities of GNR segments and location of edge states.** STM image of a (2, 1) chiral edge ($V_S = -0.97$ V, $I_t = 50$ pA). Two dashed black lines lie parallel to the edge orientation and the zigzag orientation respectively. The chiral angle between them is 19.1°. The green superimposed graphene lattice structure shows that the edge-state bright spots reside on the zigzag fragments.

**Figure S2: Edge termination of zigzag GNR.** Simulated STM images of (a) zigzag edge with one hydrogen atom per edge carbon atom and (b) zigzag edge with two hydrogen atoms per edge carbon atom (Klein edge). The images were simulated using a tight-binding Hamiltonian within the Tersoff-Hamann approximation. The bias voltage is the same as in experiments ($V_s = -0.97$ V). Solid lines correspond to covalent bonds between neighboring $sp^2$ carbon atoms, the green dots denote $sp^3$ carbon atoms. (c) Experimental image of a zigzag segment ($V_S = -0.97$ V, $I_t = 50$ pA) with superimposed lattice structure.

**Figure S3: Simulated STM images of different hydrogen terminated configurations for a (2, 1) chiral edge.** Electronically equivalent configurations shown in panels **a** and **g** are the thermodynamically most stable terminations (see Fig. 3c of the main text) and match the experiment. The images were simulated using a tight-binding Hamiltonian and the Tersoff-Hamann approximation. The bias voltage is the same as in experiments ($V_S = -0.97$ V). The solid lines correspond to covalent bonds between the neighboring $sp^2$ carbon atoms ($sp^3$-hybridized edge atoms are shown as green dots).

**Figure S4: Comparison of line profiles derived from experiment and simulation for GNR zigzag and armchair edges.** (a, b) Experimental images ($V_S = -0.97$ V, $I_t = 50$ pA) of (a) GNR zigzag and (b) GNR armchair edges, with blue regions showing areas where linescans were averaged. (c) GNR zigzag and (d) GNR armchair edge LDOS simulations with red areas indicating where linescans were averaged. Average linescan profiles for the experiment (blue lines) and the simulations (red lines) are shown for (e) GNR zigzag and (f) GNR armchair edges.

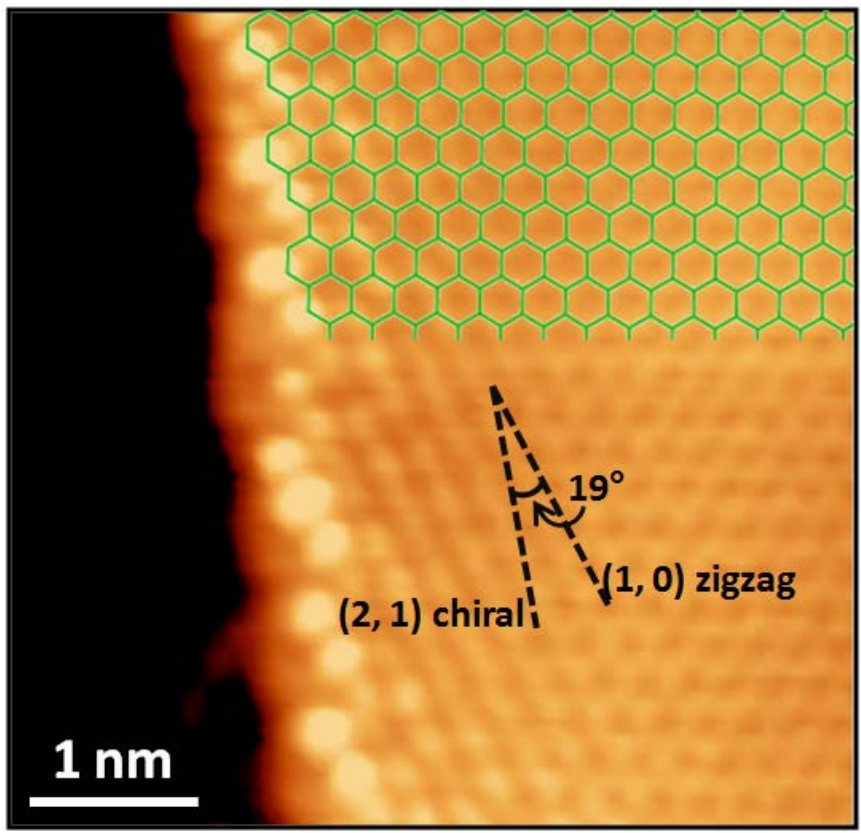

**Figure S1**

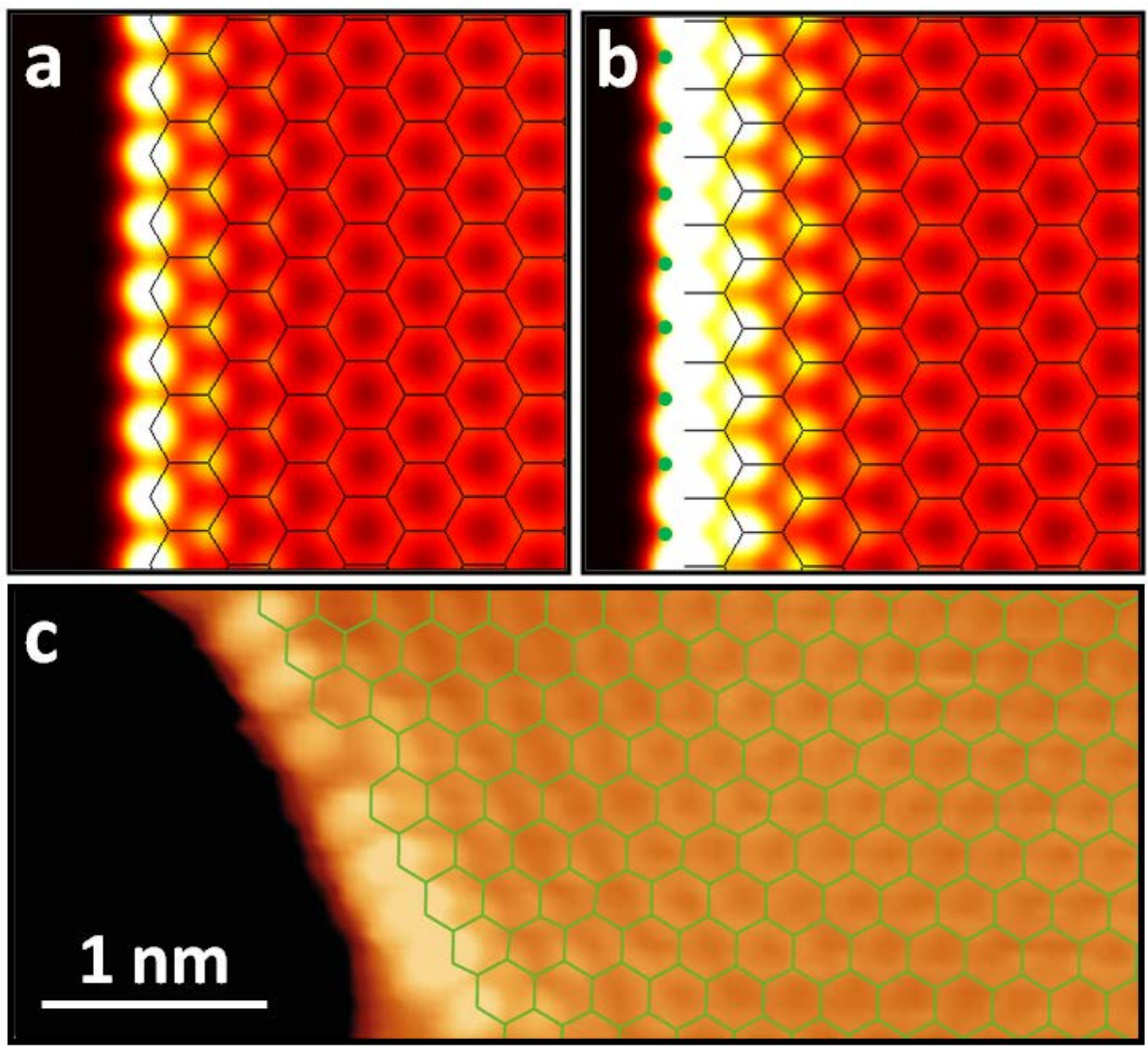

**Figure S2**

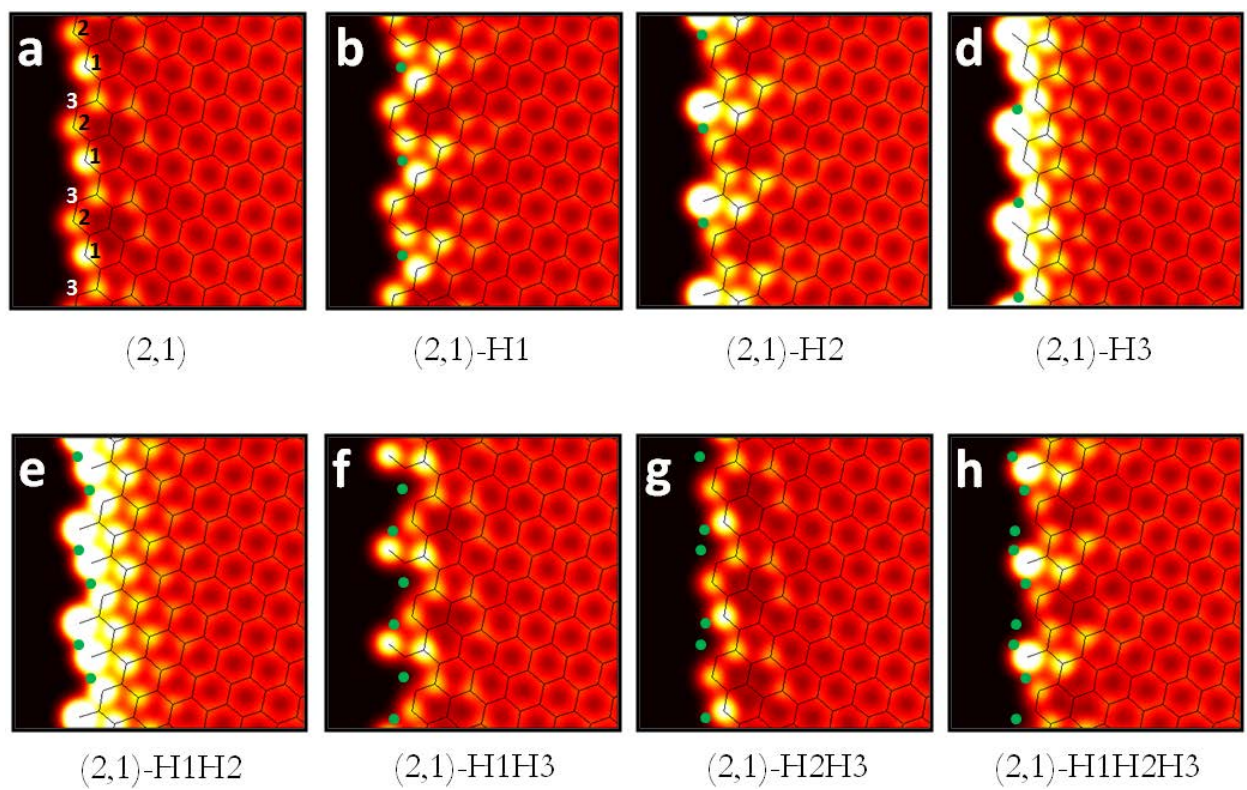

**Figure S3**

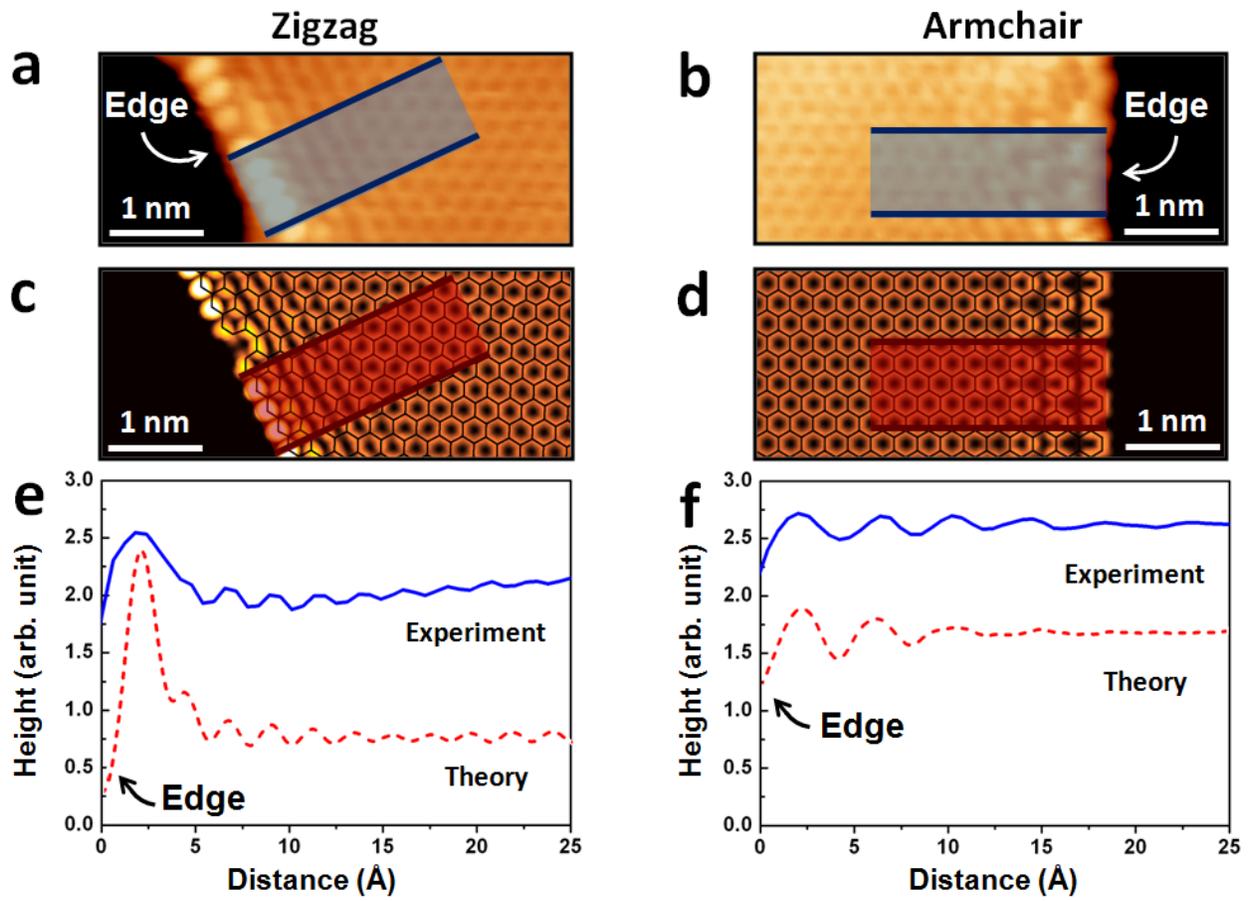

**Figure S4**